\documentclass[twocolumn,floats,aps]{revtex4}
\usepackage{mathptmx,amsmath,amsfonts,mathrsfs,psfrag,latexsym,pstricks,graphics}

\begin{document}
\title{Vortex Simulations on a 3-Sphere}
\author{O.M. Dix and R.J. Zieve}
\affiliation{Physics Department, University of California at Davis}
\begin{abstract}
We generate vortex tangles using a Hopf flow on a 3-sphere, in place of the standard torus defined by periodic boundary conditions. These tangles are highly anisotropic, with vortices tending to align along the flow direction. Standard power law dependences change accordingly from their values in more isotropic tangles. The line length density $\langle L\rangle$ is proportional to $v_{ns}^{1.28}$, where $v_{ns}$ is the drive velocity, and the reconnection rate depends roughly on $\langle L\rangle^2$. We also discuss the
effect of the full Biot-Savart law versus the local induction approximation (LIA). Under LIA the tangle collapses so that all vortices are nearly aligned with a single flow line, in sharp contrast to the torus where they become perpendicular to the driving velocity. Finally we present a few torus simulations with a helical velocity field, which in some ways resembles the 3-sphere flow.
\end{abstract}
\maketitle

\section{Introduction}

The notoriously intractable equations governing fluid flow generate a great deal
of numerical work \cite{Blazek, Maxey, Hosain, Cottet}. While many calculations deal directly with the
velocity field, others focus on the vorticity \cite{Cottet}. The latter is
natural for a variety of flows with intense vorticity, including airplane
trailing vortices, severe weather events such as hurricanes and tornadoes, and
the molecular clouds that lead to star formation. Vortex methods track the
positions of vortex filaments over time, by calculating the velocity field from
the vortex locations and then using it to update the vortices. Such calculations
are particularly well-suited to superfluid helium, where the restriction of
vorticity to angstrom-scale cores makes the ``slender filament" approximation
quite accurate, and a great deal of work has been devoted to understanding such
simulations of superfluid vortices \cite{Schwarz85, Tsubota17}.

Efforts to identify general principles necessarily focus on idealized cases, such as homogeneous turbulence. Since homogeneity is destroyed near a boundary, computations can take two approaches. One is to use a large, finite volume, but to evaluate properties only in a smaller region far from the boundaries. The other is to run calculations in a space without boundaries. The latter has several advantages, notably that there is no need to deal with the complex behavior at surfaces, and that the entire computational power can be devoted to the region of interest. The standard choice is periodic boundary conditions. Topologically, these three-dimensional simulations run on the three-dimensional torus $T^3$ rather than in infinite $\mathbb{R}^3$.

The choice of periodic boundary conditions can affect the calculation results. An example encountered in vortex filament simulations is the tendency of vortices to fall into an ``open-orbit" state, where they align parallel to each other and perpendicular to the driving velocity field \cite{Schwarz88, Aartsthesis}. In this configuration they interact only trivially with the velocity field, resulting in uniform translation of the entire set of vortices, so the open-orbit state persists indefinitely. Since this state requires the vortices to be closed loops that are perpendicular to the velocity field everywhere, it corresponds to a set of infinite straight vortex lines in $\mathbb{R}^3$. Vortex rings in $\mathbb{R}^3$, unlike those in $T^3$, cannot exhibit such behavior. The particular case of the open-orbit state can be prevented by including non-local interaction terms between vortices in addition to the Arms-Hama local term \cite{adachi10}, but it nevertheless demonstrates the possibility of topological artifacts. This raises the question of how else the topology may affect simulation results.

Here we present vortex filament simulations in a different three-dimensional space, namely a 3-sphere $S^3$, the surface of a four-dimensional ball. We find several unusual behaviors, including a tendency for vortices to align with the velocity flow instead of perpendicular to it. A particularly interesting result is the unusually high degree of directionality in homogeneous vortex tangles, which provides a testing ground for predictions about anisotropic turbulence.

\section{Computational considerations}

As is standard in vortex filament simulations \cite{Schwarz85, BarenghiSamuels04}, we consider a vortex line
$\mathbf{s}(\xi, t)$, where $\xi$ is arc length and $t$ is time. The vortex moves according to 
\begin{equation}
\mathbf{\dot{s}}= \mathbf{\dot{s}}_0 +\alpha\mathbf{
s^\prime} \times (\mathbf{ v}_n - \mathbf{ \dot{s}}_0).
\label{e:locvel}
\end{equation} 
Here $\mathbf{s^\prime}=d\mathbf{s}/d\xi$ is the unit tangent to the vortex,
$\mathbf{v}_n$ is the applied normal fluid velocity that supplies energy to the
vortex tangle, and $\alpha$ is a coefficient of mutual friction between the
normal and superfluid components. We take $\alpha=0.1$, corresponding to a
temperature of about 1.6 K in superfluid ${}^4$He. The complete version of
Equation \ref{e:locvel} includes an additional mutual friction term
$-\alpha^\prime \mathbf{s^\prime}\times(\mathbf{ s^\prime}\times
(\mathbf{v}_{ns} - \mathbf{\dot{s}}_0))$, but as is often done in helium vortex
simulations we set the coefficient $\alpha^\prime=0$ and neglect this
contribution. Finally, $\mathbf{\dot{s}}_0$ is the local superfluid velocity,
given by the Biot-Savart law. In $\mathbb{R}^3$ this is  
$$\mathbf{\dot{s}}_0 = \frac{\kappa}{4\pi}\int_\mathscr{L}
\frac{(\mathbf{s}_1-\mathbf{s})\times d\mathbf{s}_1}{|\mathbf{s}_1-\mathbf{s}|^3},$$  
where $\kappa$ is the circulation quantum and the integral is over all the vortex lines. We will discuss below the
modifications needed to this integral for use in $S^3$.

For numerical evaluation, the vortex is approximated by a sequence of points
with straight line segments connecting neighboring points. In evaluating the
velocity at $\mathbf{ s}$, the two vortex segments adjacent to $\mathbf{ s}$ must be
removed from the integral. They are replaced by a term proportional to the
local curvature of the vortex line, which is scaled to agree with the
analytically calculated velocity for circular vortex rings. The expression for
local superfluid velocity becomes 
$$\mathbf{\dot{s}}_0 = \beta \mathbf{s^\prime}\times \mathbf{s^{\prime\prime}} +\frac{\kappa}{4\pi}\int^\prime_\mathscr{L} \frac{(\mathbf{s}_1-\mathbf{s})\times d\mathbf{s}_1}{|\mathbf{s}_1-\mathbf{s}|^3},$$
where $\mathbf{s^{\prime\prime}}=d^2\mathbf{s}/d\xi^2$ is the curvature vector.
The prefactor $\beta$ is set to 
$$\beta=\frac{\kappa}{4\pi}\ln\left(\frac{2\sqrt{\ell_+\ell_-}}{e^{1/4}a_0}\right),$$
which yields the correct behavior for circular vortex loops in $\mathbb{R}^3$ with uniform point spacing.
$a_0$ is the core radius of a vortex filament, and $\ell_{\pm}$ are the distances between the point where the velocity is being calculated and its neighbors on each side. Our calculations use values
appropriate to superfluid ${}^4$He, $\kappa=9.969\times 10^{-4}$ cm$^2$/s and $a_0=1.3\times 10^{-8}$ cm.

We use a 3-sphere embedded in $\mathbb{R}^4$, consisting of points $(x,y,z,w)$
such that $x^2+y^2+z^2+w^2=R^2$. (Mathematicians often use $S^3$ for the sphere
of unit radius, but we will use this notation more generally for our sphere of radius $R$.)
Four-dimensional Cartesian coordinates help with much of  the necessary
calculation.  For the driving velocity, we use a Hopf vector field, given in
$\mathbb{R}^4$ as $\mathbf{v}=\frac{v}{R}(-y,x,-w,z)$. This Hopf velocity field
possesses several useful properties. It is tangent to $S^3$, so it drives vortex
motion entirely within the manifold. It also has uniform magnitude $v$ and zero
divergence. However, it is not irrotational; in fact its curl in $S^3$ is
$\frac{2}{R}\mathbf{ v}$, parallel to the Hopf field itself \cite{Arnold}. We
assign the driving Hopf velocity field to the normal fluid, which unlike the
superfluid is not required to be irrotational. Nonetheless, as we shall see, the
non-zero curl has consequences for the vortex behavior.

Several adjustments are needed when restricting to $S^3$. To begin with,
distances such as $|\mathbf{s}_1-\mathbf{s}|$ should be calculated along the geodesic
through the two points, rather than along the shorter chord that connects the
points in $\mathbb{R}^4$. Next, vectors defined at a point on $S^3$ lie in the
tangent space of $S^3$ at that point. Vector operations such as dot products or
cross products can be performed within a tangent space. However, in our
calculations relevant vectors are often defined at two different points of
$S^3$, where the tangent spaces themselves differ. Viewed as a vector within
$\mathbb{R}^4$, one vector may not even be in the tangent space where the other
vector is defined. For vectors in different tangent spaces, we parallel
transport one vector along a geodesic to the location of the other vector,
before carrying out any further vector operations \cite{DoCarmo}. Once we are working with two
vectors in the same tangent space, the dot product takes exactly its value in
$\mathbb{R}^4$. For cross-products in the tangent space at $\mathbf{q}$, we use the
determinant of a $4\times 4$ matrix: 
$$\mbox{cross}(\mathbf{ v_1},\mathbf{ v_2},\mathbf{
q})\cdot \mathbf{a}=\frac{1}{R}\begin{tabular}{|llll|} $v_{1x}$ & $v_{1y}$ & $v_{1z}$ &
$v_{1w}$ \\ $v_{2x}$ & $v_{2y}$ & $v_{2z}$ & $v_{2w}$ \\ $q_{x}$ & $q_{y}$ & $q_{z}$
& $q_{w}$ \\ $a_x$ & $a_y$ & $a_z$ & $a_w$ \end{tabular}$$
Here $\mathbf{a}$ is an arbitrary vector in $\mathbf{R}^4$; taking
it to be a unit vector selects the component of the cross-produce in
that direction.  Given three vectors $\mathbf{v_1}$, $\mathbf{v_2}$, and
$\mathbf{q}/R$, the cross-product is a vector orthogonal to all three and
with magnitude defined by the volume of the parallelopiped spanned by the
three vectors. Since $\mathbf{q}$ is the normal vector to the 3-sphere,
the cross-product must lie in the tangent space at $\mathbf{q}$.

Some of the vectors involved in the calculations, particularly the curvature
vectors of vortex lines as calculated in $\mathbb{R}^4$, may have components
along the 3-sphere radius vector. Such radial components give no contribute to
cross-products, but they could affect dot products. We explicitly remove any
radial component of a vector before proceeding with further operations. In the
case of curvature vectors, this has the effect of removing the intrinsic
curvature of the 3-sphere from our calculations.

Another issue is that numerical evaluation of Equation \ref{e:locvel} yields a velocity vector in a tangent space to $S^3$, not in $S^3$ itself.
Using such velocity vectors to update the locations of vortex core  would result in new locations outside the 3-sphere.
Projecting these points directly back to the 3-sphere would slightly decrease
the distance traveled during the time step. Instead, we move the vortex points
by the desired distance but along the geodesic defined by the projection of the
velocity onto $S^3$.

\begin{figure}[htb]
\begin{center}
\scalebox{.4}{\includegraphics{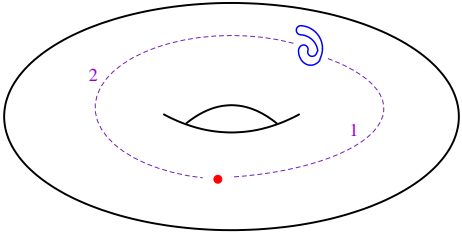}}
\caption{Two possible paths from a vortex loop (blue online) to the same point (red online) on a torus.}
\label{f:toruspaths}
\end{center}
\end{figure}

The Biot-Savart law must also be treated differently on $S^3$ from on
$T^3$. The torus has no curvature, so within regions very small
compared to the torus diameters, the Biot-Savart expression of $\mathbb{R}^3$
can be used. As the distance between the vortex segment and the test point
grows, additional contributions enter  corresponding to paths that loop around
the torus, as shown in Figure \ref{f:toruspaths} for $T^2$. The
shortest path between any part of the vortex and the point indicated lies near
the dotted line 1; in the analogous three-dimensional situation, this would give
certain Biot-Savart contributions to the velocity field at the test point. On
the other hand, the dotted line 2 corresponds to paths not much longer but in an
entirely different direction, which would give very different contributions.
Other paths, wrapping around the torus one or more times, are also possible. The
multiple contributions have sometimes been accounted for by representing the
torus as a periodic cube tiling $\mathbb{R}^3$, and adding contributions from
the original vortex segment and also from its images, out to some distance
beyond which further contributions are deemed negligible \cite{kondaurova14}.

On $S^3$ the standard Cartesian Biot-Savart expression is again merely an approximate solution, both because alternate paths may traverse the sphere in
different directions and because the curvature of the sphere makes the
resemblance to $\mathbb{R}^3$ only local. Some of our computations use the
standard Euclidean Biot-Savart law within a limited region of $S^3$. To update the position of a point $P$ along the vortex core, we need to calculate the fluid velocity at that point. We use parallel transport to map all vectors needed for the calculation into the tangent space at $P$. We then carry out the calculation within that tangent space. We calculate contributions only from vortex segments sufficiently close to $P$.

Most of our calculations use a different approach. Unlike on $T^3$,
there exists an exact form for the Biot-Savart law on $S^3$
\cite{DeTurckGluck05, Kuperberg}, which allows a more complete calculation of the velocity
field produced by vortex segments. We generally use the full 3-sphere
Biot-Savart law. We retain contributions to velocity either from vortices on the
entire sphere, or from those in a limited region surrounding the evaluation point. Whether we use the Cartesian Biot-Savart law or the exact expression on $S^3$, we project the calculated velocity vector onto a geodesic and move $P$ an appropriate distance along that geodesic.

We follow the presentation by DeTurck and Gluck of the Biot-Savart law on the 3-sphere \cite{DeTurckGluck05}. The contribution to the
velocity field at a point $\mathbf{q}$ from a vortex line $U$ on $S^3$ is given by
\begin{equation}
d\mathbf{v}=\int_U \nabla_{q} \phi(u,q) \times P_{qu}d\mathbf{u}.
\label{e:biotint}
\end{equation} 
Here $P_{qu}$ is the
parallel transport operation described above, which takes a vector from the tangent space at $\mathbf{u}$ to the tangent space at $\mathbf{q}$. The function
$\phi$ depends only on the
distance $h=|\mathbf{q}-\mathbf{u}|$ along the geodesic connecting its two arguments and has the form
\begin{equation}
\phi(h)=\frac{\kappa}{4\pi^2R^2}\left(\pi-\frac{h}{R}\right)
\csc\frac{h}{R}.
\label{e:phidef}
\end{equation}
The familiar Biot-Savart law in $\mathbb{R}^3$ would instead use
$\phi(h)=\kappa/4\pi h$, with $h$ the distance along the straight-line path
from $\mathbf{q}$ to $\mathbf{u}$.

\begin{figure}[htb]
\begin{center}
\scalebox{.4}{\includegraphics{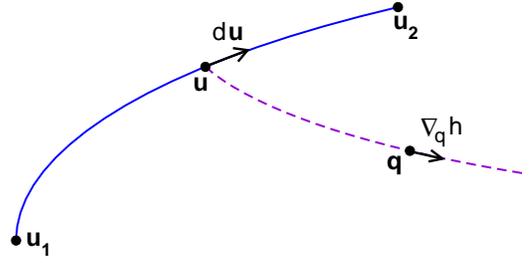}}
\caption{Integration along a geodesic (solid curve), to evaluate Biot-Savart contribution at point $\mathbf{q}$. The dashed curve is the geodesic between $\mathbf{q}$ and a point $\mathbf{u}$ on the integration path.}
\label{f:biotpic}
\end{center}
\end{figure}

For the Biot-Savart calculation we consider the vortex core as a series of
points connected by straight segments, i.e. geodesics. We evaluate the
integral of Equation \ref{e:biotint} exactly along each of these finite-length geodesic segments, as shown in
Figure \ref{f:biotpic}. In the integrand, $\nabla\phi$ is calculated in the
tangent space at $\mathbf{q}$. Since $\phi$ depends only on the
separation between the points $\mathbf{u}$ and $\mathbf{q}$, the gradient must be in the direction where this separation changes the fastest,
which is the direction of the geodesic containing the two points. We use
$$\nabla_q \phi = \frac{\partial \phi}{\partial h}\nabla_q h$$ 
and observe that $\nabla_qh$ has magnitude 1 and is directed away from $\mathbf{u}$ along the geodesic connecting $\mathbf{u}$ and $\mathbf{q}$. 

Next consider the effect of parallel transport along this geodesic. The geodesic
lies in a plane of $\mathbb{R}^4$, defined by the tangent to the geodesic and
the normal to the 3-sphere at any point on the geodesic, such as $\mathbf{u}$.
Parallel transport of any vector rotates the vector within this plane, without
affecting the components orthogonal to the plane. This has two implications for
the calculation at hand. First, no parallel transport calculation is necessary
for the integral of Equation \ref{e:biotint}. Only the components of
$P_{qu}d\mathbf{u}$ within the tangent space but perpendicular to the geodesic
contribute to the cross-product in the integrand, and these are exactly the
components unchanged by parallel transport. Second, we can use $\mathbf{u}$ to construct a unit vector along the geodesic.
The plane of a geodesic contains the normal vectors to $S^3$ at every point on that geodesic, so
$\mathbf{u}$ can be decomposed into one component along $\mathbf{q}$ and another
tangent to the geodesic at $\mathbf{q}$. Its projection into the tangent space
at $\mathbf{q}$ is this latter component. Since the angle between $\mathbf{u}$ and $\mathbf{q}$ is $h/R$, the two components have magnitude $R\cos \frac{h}{R}$ and
$R\sin \frac hR$. Rescaling, the projection
of $-\frac{1}{\sin(h/R)}\mathbf{\hat{u}}$ into the tangent space at $\mathbf{q}$
is a unit vector along the geodesic, directed away from $\mathbf{ u}$. For Equation \ref{e:biotint} we can use $-\frac{1}{\sin(h/R)}\mathbf{\hat{u}}$ rather than $\nabla_qh$; although these vectors are not identical, the component of the former
perpendicular to the tangent space does not contribute to the cross-product. Using this fact and explicitly calculating $\partial \phi/\partial h$, Equation \ref{e:biotint} becomes
$$d\mathbf{v}=\frac{\kappa}{4\pi^2R^2}\int_{\mathbf{u}_1}^{\mathbf{u}_2} d\mathbf{u}\times \mathbf{\hat{u}}\frac{\sin\frac{h}{R}
+(\pi-\frac{h}{R})\cos\frac{h}{R}}{\sin^3\frac{h}{R}}.$$

Up to this point the discussion does not depend on the shape of the integration
path from $\mathbf{u}_1$ to $\mathbf{u}_2$. We now take that path to be a
geodesic. This is distinct from the geodesic of the previous paragraph, which
goes from the test point to the integration path. Next we set coordinates for
evaluating the integral. The geodesic integration path defines a plane in
$\mathbb{R}^4$, which we take as the $xy$-plane. We take the projections of
$\mathbf{q}$ within and perpendicular to the $xy$-plane to lie along the
positive $x$ and $z$ axes, respectively. Thus $\mathbf{ u}=R(\cos
\theta,\sin\theta,0,0)$, with $\theta_1$ and $\theta_2$ denoting the integration
limits, and $\mathbf{q}=R(\cos\psi,0,\sin\psi,0)$, where $0\le\psi\le \pi/2$.
Hence $d\mathbf{ u}=R(-\sin\theta,\cos\theta,0,0)d\theta$ and the integral
becomes
$$d\mathbf{v}=-\frac{\kappa}{4\pi^2}\mathbf{\hat{w}}\int_{\theta_1}^{\theta_2} d{\theta}\sin\psi
\frac{\sin\frac{h}{R}
+(\pi-\frac{h}{R})\cos\frac{h}{R}}{\sin^3\frac{h}{R}},$$ 
where $h$ depends on $\theta$. We can also adjust the integral so that
$\sin\theta\geq 0$, as follows. If the original choice of coordinates gives
$\sin\theta\leq 0$ along the entire geodesic, then rotation by $\pi$ in the
$yw$-plane switches the geodesic segment to positive $\theta$ without altering
$\mathbf{q}$. If $\sin\theta$ changes sign along the geodesic, then the curve
can be divided into two segments and the calculation carried out separately for
the two. The geodesic segments we integrate along connect consecutive points on a vortex core and are always too short for $\sin\theta$ to change sign more than once.

The variables $\theta$, $\psi$, and $h$ are related through 
$$\cos\frac{h}{R}=\frac{\mathbf{ u}\cdot\mathbf{ q}}{R^2} =\cos\theta \cos\psi.$$
Using $-\frac{1}{R}\sin\frac{h}{R}dh=-\sin\theta\cos\psi d\theta$ and
assuming $\sin\theta\geq 0$, we have 
$$d\mathbf{v}=-\frac{\kappa\sin\psi}{4\pi^2R}\mathbf{\hat{w}}\int_{h_1}^{h_2}dh
\frac{\sin\frac{h}{R}
+(\pi-\frac{h}{R})\cos\frac{h}{R}}{\sin^2\frac{h}{R}\sqrt{\cos^2\psi -
\cos^2\frac{h}{R}}}.$$ 
An additional minus sign would appear if $\sin\theta \cos\psi < 0$, but our choice of coordinates ensures that this is not the case. Integrating gives
$$d\mathbf{v}=\mathbf{\hat{w}}\frac{\kappa\sin\psi}{4\pi^2(\cos^2\psi - 1)} \bigg(-\arcsin \left(\frac{\cos \frac hR}{\cos\psi}\right) $$ 
$$+\frac{(\pi-\frac{h}{R})\sqrt{\cos^2\psi -
\cos^2\frac{h}{R}}}{\sin\frac{h}{R}}\bigg)\bigg|_{h_1}^{h_2}$$
where our coordinates again eliminate any ambiguity in the sign of the first term. In the coordinates used above for the geodesic segment, the contribution is always in the $\mathbf{w}$ direction. We then rotate it back into the original coordinate frame in which the vortices are defined, repeat for each geodesic segment along the vortices, and add the results to obtain the Biot-Savart integral along the entire set of vortices.

\begin{figure}[htb]
\begin{center}
        \begin{tabular}{l l}
        (a) & (b) \vspace{-.2in}\\
	\hspace*{-.3in}
        \scalebox{.4}{\includegraphics{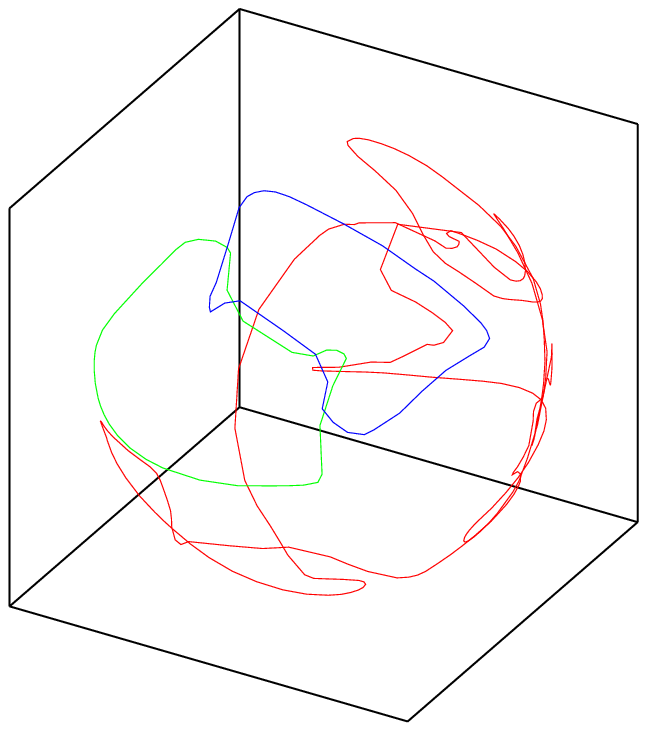}}
        & 
	\hspace*{-.3in}
        \scalebox{.4}{\includegraphics{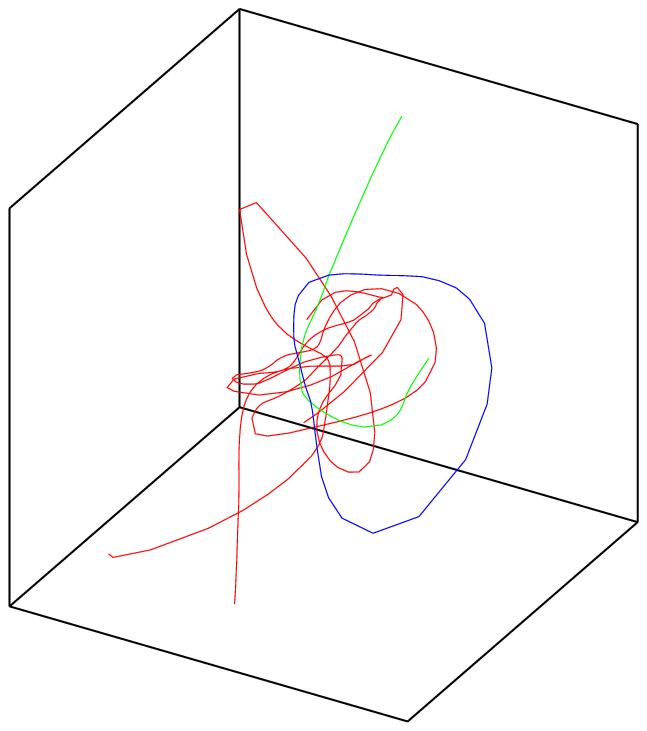}}\\
        (c) &(d) \vspace{-.2in}\\
	\hspace*{-.3in}
        \scalebox{.4}{\includegraphics{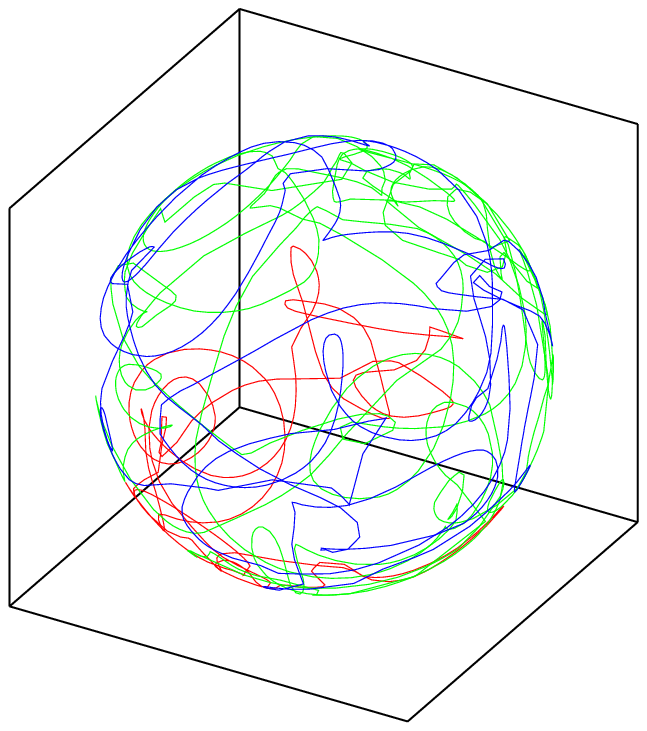}}
        & 
	\hspace*{-.3in}
        \scalebox{.4}{\includegraphics{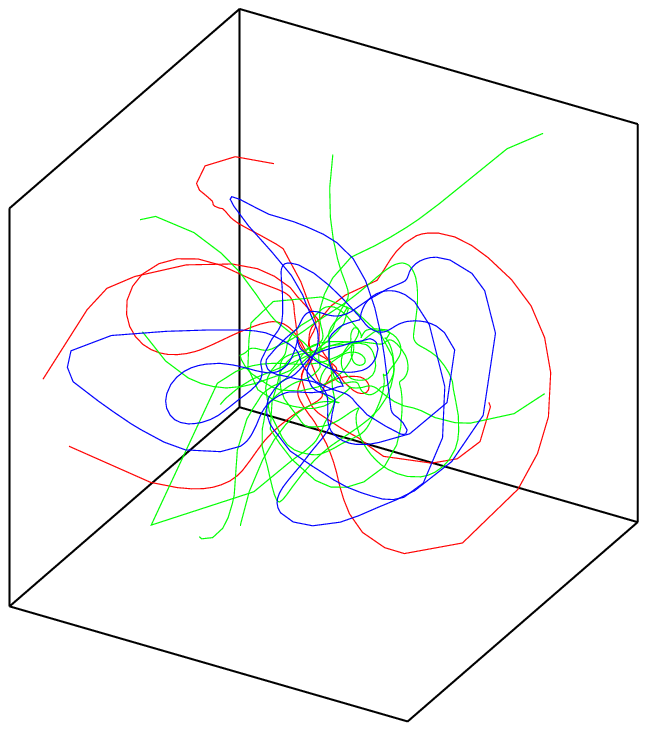}}\\
        (e) & (f) \vspace{-.2in}\\
 	\hspace*{-.3in}
        \scalebox{.4}{\includegraphics{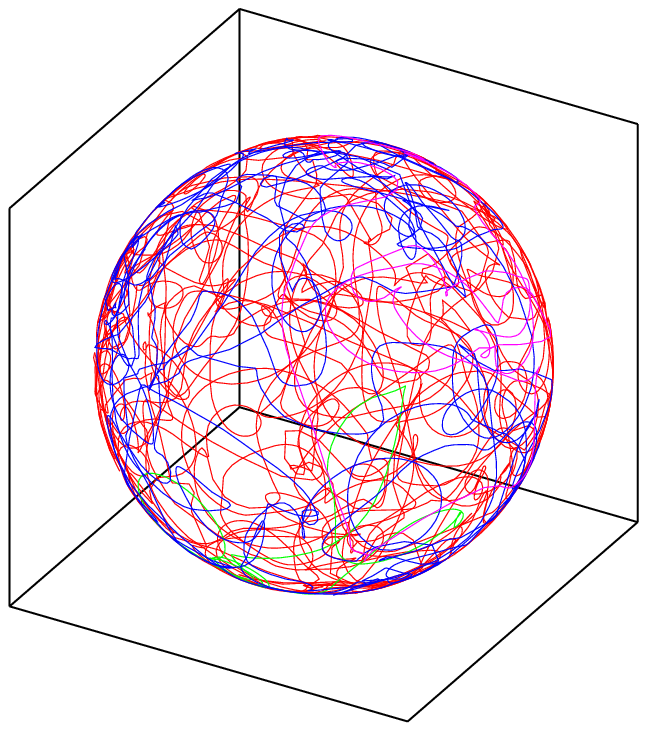}}
        &
	\hspace*{-.3in}
        \scalebox{.4}{\includegraphics{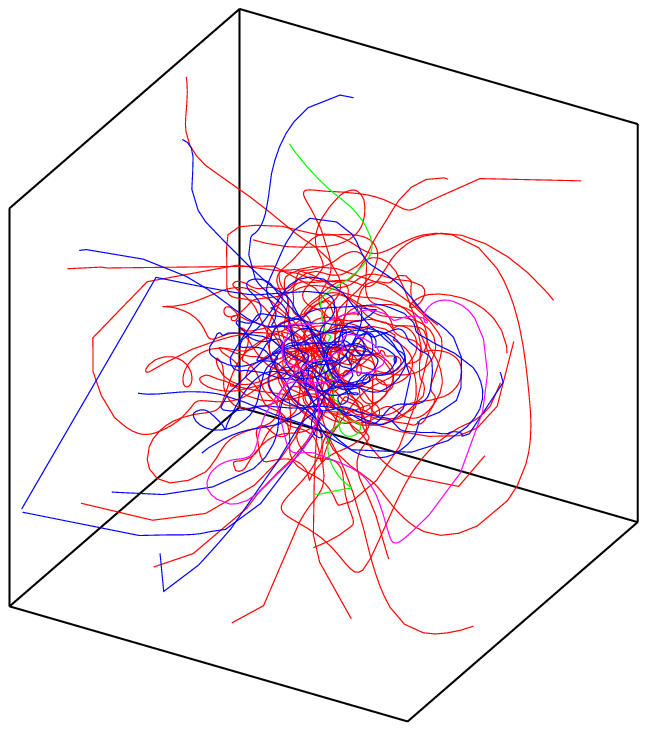}}
        \end{tabular}
\caption{Hopf and stereographic projections of vortex tangles for $v_{ns}$ = 1, 2, 4 cm/s.}
\label{f:tangles}
\end{center}
\end{figure}
 
\section{Properties of stable tangles}

\begin{figure}[htb]
\begin{center}
\scalebox{.45}{\includegraphics{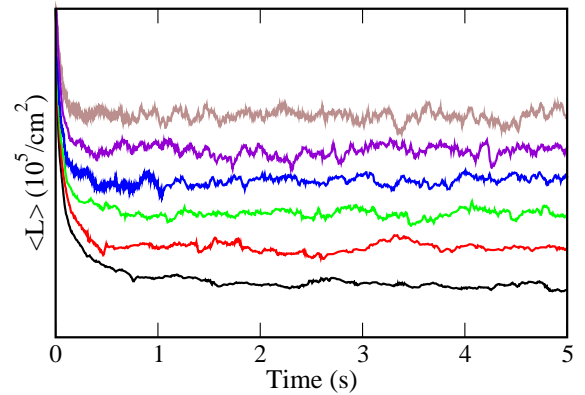}}
\caption{Line length density as a function of time, for (bottom to top) $v_{ns}$ = 1, 1.2, 1.4, 1.6. 1.8, 2 cm/s.}
\label{f:timedevel}
\end{center}
\end{figure}

Calculations using these equations successfully produce vortex tangles,
as shown in Figure \ref{f:tangles}.  We use two types of projection, Hopf
and stereographic, to
display the vortex configuration. These are described in more detail in
the appendix, as well as in algebraic topology textbooks \cite{thurston}. The right column of Figure \ref{f:tangles} uses
stereographic projection, which maps $S^3$ onto flat three-dimensional
space $\mathbb{R}^3$. One point of $S^3$ goes to infinity, and its
antipode maps to the origin. A sphere within $S^3$, centered at either
of these points, maps to a sphere in $\mathbb{R}^3$ centered at the
origin. Hence the half of $S^3$ closest to the point sent to the origin
maps to a ball centered at the origin. The other half of $S^3$ maps to
the remainder of $\mathbb{R}^3$ outside of the ball. Some portions of the
vortex lines in Figure \ref{f:tangles} lie outside the displayed region.
The second type of visualization, shown in the left column of Figure
\ref{f:tangles}, is the Hopf projection, which maps $S^3$ onto $S^2$, the
two-dimensional surface of a ball in $\mathbb{R}^3$. Each
Hopf fiber maps to a single point of $S^2$. This mapping is particularly
informative when vortices align closely with the velocity field, since
after projection such vortices appear as points or very small loops.

As seen previously in many calculations using periodic boundary
conditions \cite{Schwarz88, Aartsthesis, adachi10, kondaurova14, baggaley12, owentorus}, the vortex tangle eventually reaches a steady state
where various statistical properties remain constant over time. Figure
\ref{f:timedevel} shows the time development of the average vortex line
density, for several different driving velocities. The flat portion of
each curve corresponds to the steady-state situation. As the driving
velocity $v_{ns}$ increases, the line density increases monotonically
and the time to reach the steady-state density decreases. 

\begin{figure}[b]
\begin{center}
\hspace*{-.15in}
\scalebox{.45}{\includegraphics{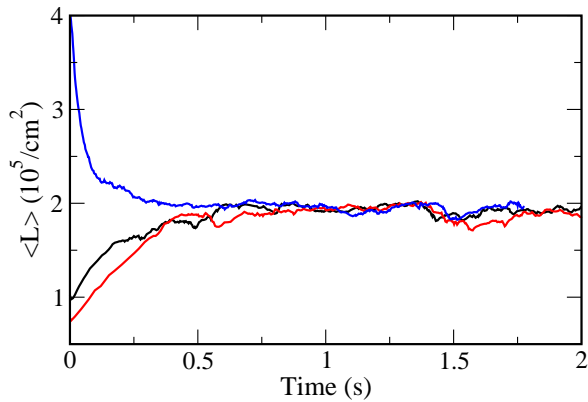}}
\caption{Line length density $\langle L\rangle$ as a function of time, for $v_{ns} = 1.4$ cm/s and non-local interactions calculated within a
hemisphere, for three different initial configurations. The two calculations with initially increasing $\langle L\rangle$ use the exact
Biot-Savart law, while the third uses the approximate Biot-Savart formula.}
\label{f:initial}
\end{center}
\end{figure}

We find that the statistical behavior of tangles is independent of the
initial conditions for the calculations, as shown in Figure \ref{f:initial}.
Furthermore, the more familiar, approximate Biot-Savart calculation provides
confirmation that the exact formula functions correctly. The two agree well for
dynamics of simple vortex configurations, such as a single circular vortex ring.
Their results for line length density and tangle anisotropy also agree, not only
for very short-distance calculations of vortex interactions, where the
approximation should be quite good, but even when applied on as much as a full
hemisphere. Thus Figure \ref{f:initial} shows that line length density
equilibrates at the same level for three simulations with different initial
conditions, two using the exact Biot-Savart law and the third using the
approximate version. In each case vortex interactions out to a hemisphere were
included. We do not attempt to use the approximate Biot-Savart law at larger
distances.

\begin{figure}[htb]
\begin{center}
\hspace*{-.15in}
\scalebox{.45}{\includegraphics{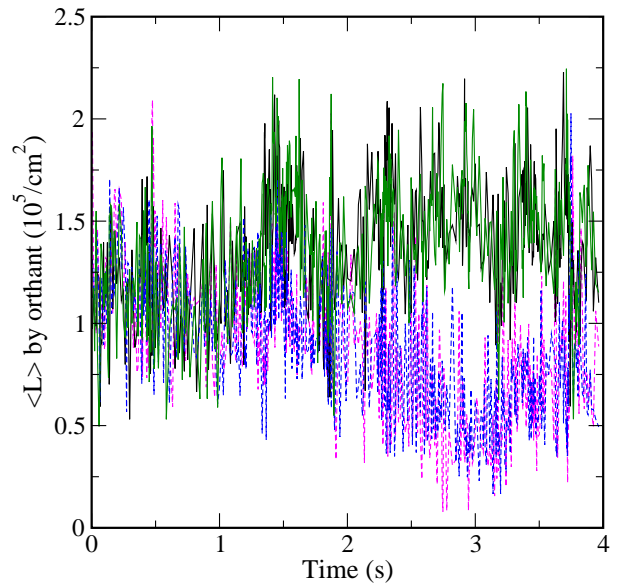}}
\caption{Line length density as a function of time for $v_{ns}$ = 1
cm/s, for orthants 9 (magenta, dashed), 10 (black, solid), 12 (blue,
dashed), and 15 (green, solid). The two solid curves track each other
closely, as do the two dashed curves, showing rapid homogenization in
the corresponding orthants.}
\label{f:homogsep}
\end{center}
\end{figure}

Despite the rapid achievement of a steady-state line density in Figure
\ref{f:timedevel}, the tangles, particularly for low $v_{ns}$, are not
entirely homogeneous. Figure \ref{f:homogsep} shows the line density
in four of the sixteen orthants of $S^3$. Although the line density
is very similar in certain pairs of orthants, in others it differs
significantly. We number the orthants by assigning one bit to each of
the four Cartesian coordinates of the space $\mathbb{R}^4$ that contains
the 3-sphere. A negative coordinate corresponds to a bit value of zero, and a
positive or zero coordinate to a value of one.  We place the $x$ bit in the
rightmost position, then the $y$ bit, then $z$, and finally the $w$ bit in the
leftmost position. Thus orthant 1 has $x\geq 0$, $y<0$, $z<0$, and $w<0$, while
orthant 13 has $x\geq 0$, $y<0$, $z\geq 0$, and $w\geq 0$. As Figure
\ref{f:homlongtime} shows, the line densities in certain pairs of orthants
continue to diverge for several seconds, even though the long-time average line
density is the same in both orthants. The issue is not a slow approach to the
steady state line density, but rather slow communication between different
regions of $S^3$ even in the steady state.

\begin{figure}[htb]
\begin{center}
\hspace*{-.15in}
\scalebox{.45}{\includegraphics{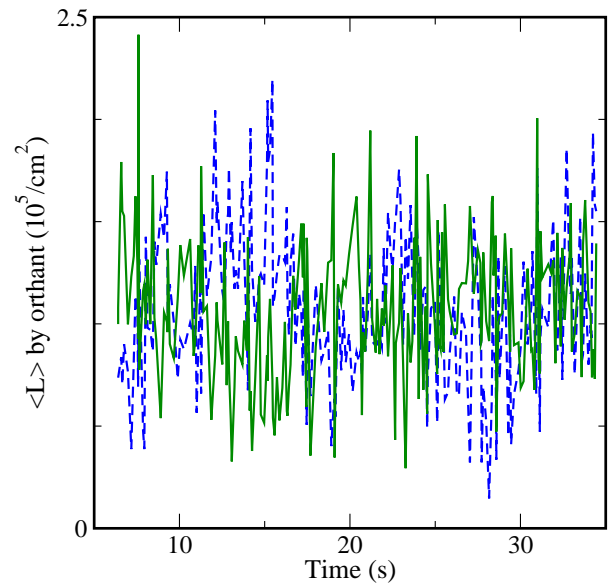}}
\caption{Continuation of line length density from Figure \ref{f:homogsep}, for orthants 12 (blue, dashed) and
15 (green, solid). Over sufficiently long times the average density is the same in each orthant.}
\label{f:homlongtime}
\end{center}
\end{figure}

\begin{figure}[htb]
\begin{center}
\hspace*{-.15in}
\scalebox{.6}{\includegraphics{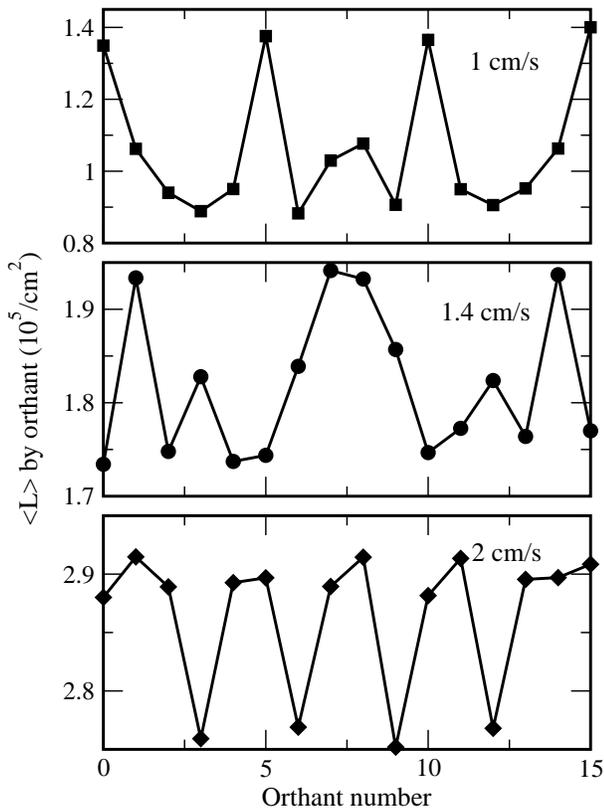}}
\caption{Line length density for the sixteen orthants of $S^3$, for
steady-state tangles with $v_{ns}=1$ cm/s (top), $v_{ns}=1.4$ cm/s
(center), and $v_{ns}=2$ cm/s (bottom). In each case the average is
over a 4-second interval.} 
\label{f:homorthant}
\end{center}
\end{figure}

The top frame of Figure \ref{f:homorthant} shows the line length density for all
sixteen orthants, averaged over the same four seconds of steady-state tangle
shown in Figure \ref{f:homogsep}.  The orthants separate into four groups by
line density. Orthants 0, 5, 10, and 15 have by far the largest line density. 
The next group is orthants 1, 7,  8, and 14; then orthants 2, 4, 11, and 13 with
nearly identical line length densities; and finally orthants 3, 6, 9, and 12.
These groupings track the velocity field. For example, consider the point
$(-\frac12, -\frac12,-\frac12,-\frac12)$, which lies exactly in the center of
orthant 0. The flow line through this point also passes through quadrants 5, 10,
and 15; in fact, it reaches the points $(\frac12,-\frac12,\frac12,-\frac12)$,
$(\frac12,\frac12,\frac12,\frac12)$ and $(-\frac12,\frac12,-\frac12,\frac12)$ at
the centers of these orthants. Likewise the centers of orthants 1, 7, 8, and 14
lie on a common flow line, and similarly for the remaining two sets of orthants.
A generic flow line actually passes through {\em eight} orthants, two of the
listed sets of four.  The portion of the flow line within an orthant is always
identical for the four orthants in a set. In particular, flow lines which pass
through a large portion of one orthant do the same for the other orthants in the
same group, alternating with briefer sections in the orthants of a second group.
Hence it makes sense to think of the velocity field as roughly conveying any
vortex tangle from one orthant into the other orthants grouped with it. The
implication for homogeneity is that when vortices are swept directly from one
orthant to another through the influence of the applied velocity field, the line
length in those orthants quickly equilibrates. The process is much slower for
orthants in different groups, which do not benefit from this convective
mechanism. The resulting lack of global homogeneity for tangles on $S^3$ differs
from the situation on $T^3$, where vortex tangles remain homogeneous.
As we will see shortly, the reason may be the different anisotropy of tangles in
the two spaces.

The same presentation for higher driving velocities appears in the bottom two
frames of Figure \ref{f:homorthant}. The same groupings of orthants appear,
since that depends solely on the geometry of $S^3$. However, at each velocity
different sets of orthants have the largest and smallest line densities. As the
velocity increases, the tangle becomes more homogeneous, with variations in
$\langle L\rangle$ decreasing in both relative and absolute terms. At $v_{ns}=2$
cm/s, three sets of orthants have indistinguishable line densities, with the
variations among orthants within a set larger than any differences in line
density between sets. The fourth set of orthants has a
smaller line density than the other three, but only by about 5\%.

\begin{figure}[htb]
\begin{center}
\hspace*{-.15in}
\scalebox{.35}{\includegraphics{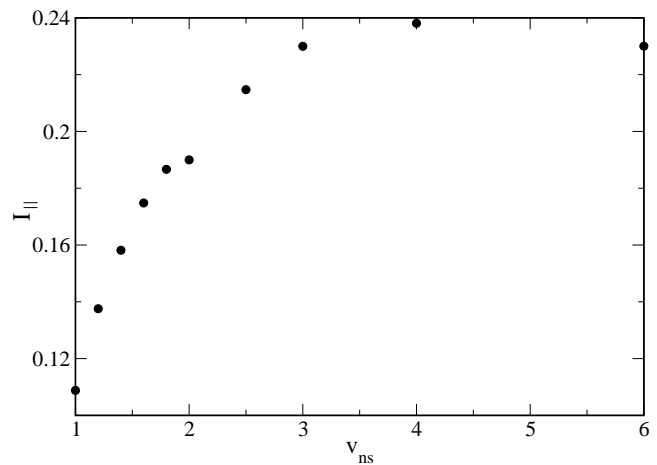}}
\caption{Anisotropy parameter $I_\parallel$ as a function of applied velocity. For
an isotropic tangle, $I_\parallel=2/3$.}
\label{f:fullanisot}
\end{center}
\end{figure}

In simulations on $T^3$, vortices have a slight tendency to align
perpendicular to the applied velocity field. We use the anisotropy parameter
\cite{Schwarz88}
$$I_\parallel=\frac{1}{VL}\int[1-(\mathbf{ \hat{s}^\prime}\cdot\mathbf{\hat{r}_\parallel})^2] d\xi$$
to quantify this effect. Here $\mathbf{\hat{r}_\parallel}$ is a unit vector
parallel to the applied flow, $VL$ is the total length of vortices, and the
integral is taken over all vortex lines. If the vortices were entirely aligned
with the applied velocity field, then $I_\parallel$ would vanish. In the other
extreme, with vortices always perpendicular to the applied field,
$I_\parallel=1$. In a perfectly isotopic tangle $I_\parallel=\frac23$. Our
previous calculations on $T^3$ give $I_\parallel=0.76$ \cite{owentorus}. Other simulations
\cite{Schwarz88, Tsubota00, adachi10, kondaurova14} find similar values for
$I_\parallel$ at $\alpha=0.1$, with a decrease towards 2/3 as $\alpha$
decreases. As shown in Figure \ref{f:fullanisot}, the vortex orientation within
the 3-sphere tangles is dramatically different. $I_\parallel$ lies between 0.1
and 0.25, depending on the driving velocity. At sufficiently high velocity the
anisotropy levels off at about 0.23, implying a significant alignment of
vortices with the applied velocity.

Alignment of vortices along flow lines limits the effectiveness of convection
at homogenizing a tangle. Vortices move roughly parallel to their own
direction, and the flow does not force them into interaction with other
vortices. By contrast, a vortex locally perpendicular to the velocity field
sweeps through a region of space, making vortex crossings more likely. Such
a vortex also drags its more distant portions along with it, increasing
communication among all parts of the tangle.

The misalignment of the vortices remains large enough to sustain vortex-vortex
interactions. Thus polarized turbulence arises naturally on $S^3$ with a Hopf
driving field. The structure is entirely different from that of counterflow
turbulence on $T^3$, where the polarization is far less and is
directed perpendicular to the flow. Anisotropic turbulence in physical
experiments can arise from rotation \cite{Swanson83}, the geometry of the
container boundaries, or entrainment by normal fluid turbulence
\cite{Barenghi02}. Boundaries and a turbulent normal fluid are both
computationally expensive, but simulations on $S^3$ provide another method of achieving highly polarized turbulence.

\begin{figure}[htb]
\begin{center}
\hspace*{-.15in}
\scalebox{.45}{\includegraphics{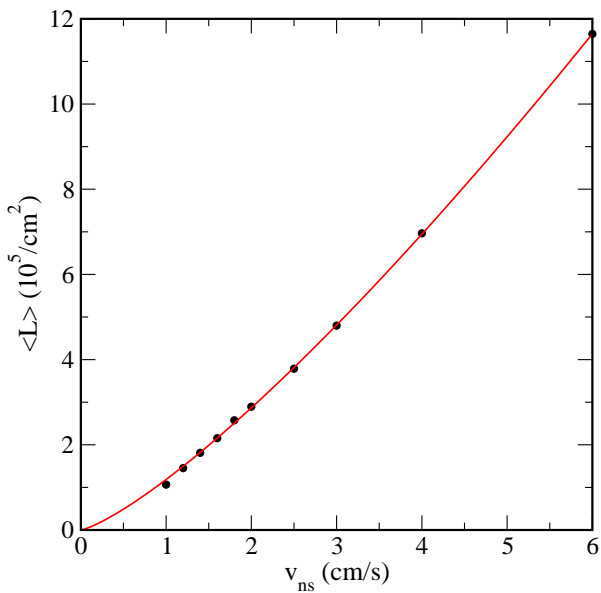}}
\caption{Line length density as a function of applied velocity. Solid line is a
two-parameter fit, yielding $1.19v_{ns}^{1.28}$, where only velocities
$v_{ns}\geq 2$ are used for the fit.}
\label{f:lvsvfit}
\end{center}
\end{figure}

One illustration of the difference between isotropic and anisotropic turbulence
is the dependence of the vortex line density on the applied velocity field.
Scaling arguments \cite{Schwarz88, SwansonDonnelly85} give a clear prediction
for isotropic homogeneous tangles: $\beta\langle L\rangle^{1/2}$ should be
proportional to $v_{ns}$. This has been observed in experiment \cite{martin83}
and simulations \cite{Schwarz88, adachi10, kondaurova14}.
The scaling breaks down when rotation introduces anisotropy \cite{Swanson83, Tsubota03}. In these studies the anisotropy parameter changes with $v_{ns}$, obscuring the dependence of $\langle L\rangle$ on $v_{ns}$. For our system,
Figure \ref{f:lvsvfit} shows the line length after
equilibration at each velocity, along with a power-law fit. Rather than finding
that $\langle L\rangle$ goes roughly as $v_{ns}^2$, we find an exponent of only
1.28. The reduced exponent is not caused by the lack of global homogeneity.
For the fit shown in Figure \ref{f:lvsvfit} we use only those trials with
velocity at least 2 cm/s, which are nearly homogeneous. Extrapolating the fit
curve to lower velocities gives excellent agreement with the line densities
found at these low driving velocities, and the fit itself changes little when
these points are included. Thus we attribute the change in exponent to the
anisotropy of the tangle; certainly anisotropy invalidates the scaling arguments
used to derive the quadratic dependence of line density on $v_{ns}$ in isotropic
tangles. In fact a reduced exponent makes intuitive sense. Vortices perfectly
aligned with the driving velocity cannot gain energy from the applied fields;
the $\mathbf{\hat{s}^\prime} \times \mathbf{v_{ns}}$ part of the friction term
vanishes. Line length increase can only originate from non-aligned vortices, and
since these are underrepresented in our tangles, our line length is less than in
the isotropic case. 

\begin{figure}[htb]
\begin{center}
\hspace*{-.15in}
\scalebox{.43}{\includegraphics{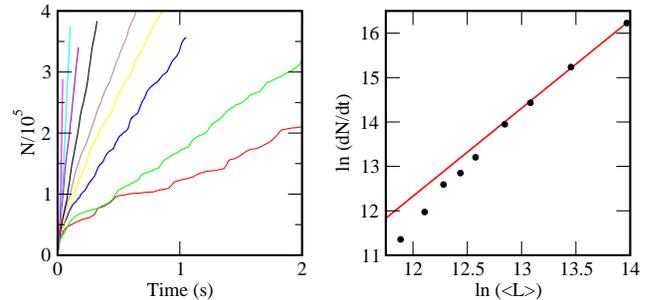}}
\caption{Left: number of reconnections $N$ as a function of time, for several
different velocities. Right: relation between reconnection rate and line length
density. Each point corresponds to one of the curves in the left graph, with
$dN/dt$ obtained from a linear fit to the long-time part of the $N(t)$ curve.
The solid line represents a power-law fit, $dN/dt = c\langle L\rangle^m$, to
the highest four points, selected because the corresponding tangles have
similar anisotropy. The resulting exponent is $m=1.98$.} 
\label{f:reconrate}
\end{center}
\end{figure}

The rate of vortex reconnections can also depend on the structure of the
tangle. For isotropic homogeneous turbulence, simple arguments \cite{Poole03,
Nemirovskii06} show that the reconnection rate is related to the vortex line
density through $dN/dt\propto \langle L\rangle^{5/2}$, which agrees with
simulations \cite{Tsubota00, Poole03}. Polarization should reduce the number
of reconnections \cite{Lvov07}, since aligned vortices encounter each other
less often. Simulations show a reduction of about a factor of two in the number
of reconnections in polarized turbulence \cite{Laurie15a}; here the polarization
comes from a normal fluid driving velocity taken from a snapshot of classical
turbulence. Polarization is also expected to decrease the scaling exponent.
In calculations of counterflow turbulence
before the vortex line length stabilized \cite{BarenghiSamuels04}, increasing
vortex line length produced both an increase in polarization and a decrease in
the reconnection rate exponent. Tsubota et al. also found a hint of a reduced
exponent \cite{Tsubota04}, albeit without enough data to identify the
exponent. In Figure \ref{f:reconrate} we fit reconnection rate vs. velocity,
using only the points corresponding to the four largest driving velocities,
where the anisotropy parameter is changing little. We find that the exponent is
indeed reduced, to 1.98. The resulting fit does not extrapolate well to the
data from lower driving velocity, but that is not surprising since the
low-velocity tangles are much more anisotropic, as seen in Figure \ref{f:fullanisot}.

\begin{figure}[htb]
\begin{center}
\hspace*{-.15in}
\scalebox{.45}{\includegraphics{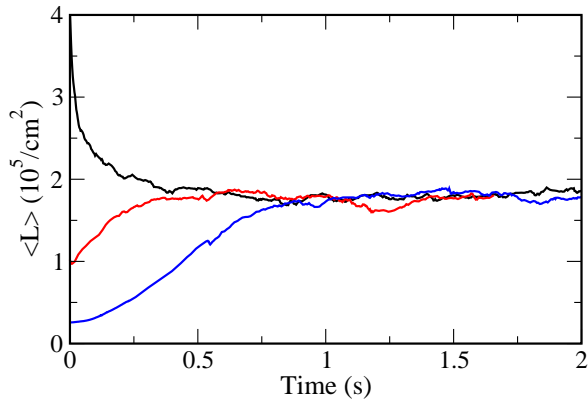}}
\caption{Line length density as a function of time, for $v_{ns} = 1.4$ cm/s and non-local interactions over the entire volume, for three different initial configurations.}
\label{f:initialfull}
\end{center}
\end{figure}

Both the average vortex line density and the tangle anisotropy are independent
of the initial configuration used. For example, Figure \ref{f:initialfull}
shows that line density for $v_{ns}=1.4$ cm/s ultimately reaches the same
stable level for three initial conditions. The convergence to a common global
line density happens despite the inhomogeneity among orthants. This
demonstrates that equilibration occurs even with weak direct communication
among different parts of the space.

\section{Local Induction Approximation}

A calculational issue on the 3-torus has been the reliability of the local
induction approximation (LIA) that ignores the non-local integral in the
equation of motion. Without the non-local term, the tangle can degenerate into
an ``open-orbit" state, consisting of straight parallel vortex lines that do not
interact \cite{Schwarz88, Aartsthesis}. Keeping the non-local term solves this
problem \cite{adachi10}, but given the extra computational demands of the full
integral even subsequent work has used LIA \cite{salman13, gorder13}, compared
results from LIA and the full integral \cite{kondaurova14, laurie15b}, or drawn
conclusions from prior LIA work \cite{nemirovskii19}. Hence better understanding
its limitations remains relevant.

\begin{figure}[htb]
\begin{center}
        \begin{tabular}{l l}
        (a) \\
	\multicolumn{2}{c}{
 	\hspace*{-.4in}
   \scalebox{.45}{\includegraphics{collapseanisot.eps}}}\\
        (b) & (c) \vspace{-.2in}\\
	\hspace*{-.3in}
        \scalebox{.4}{\includegraphics{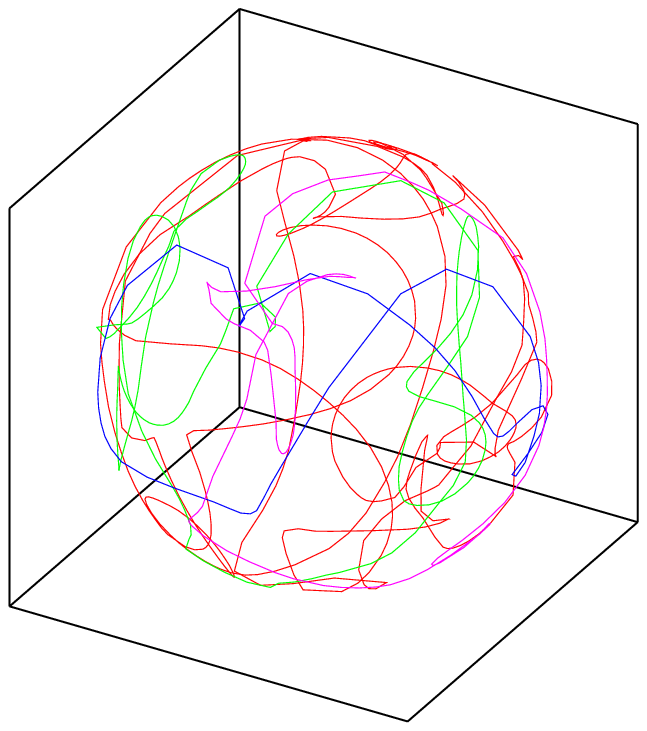}}
        & 
	\hspace*{-.3in}
        \scalebox{.4}{\includegraphics{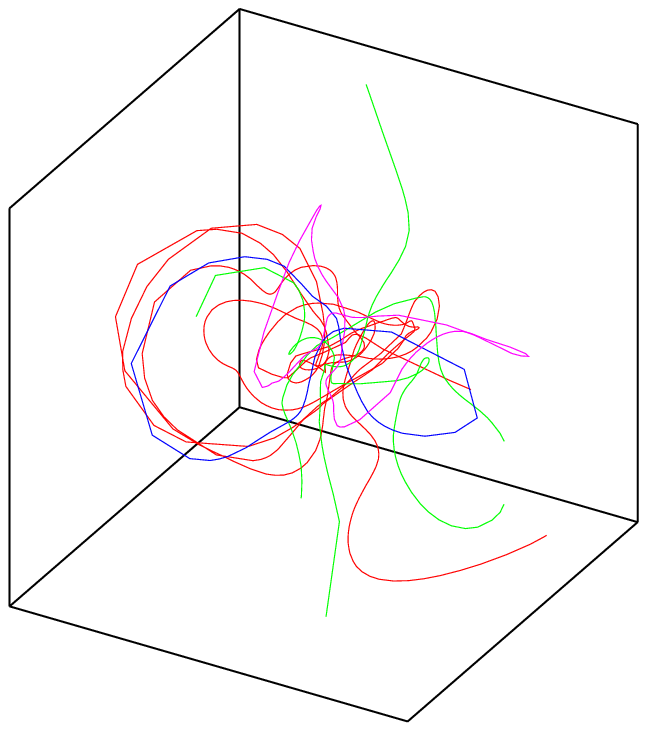}}\\
        (d) &(e) \vspace{-.2in}\\
	\hspace*{-.3in}
        \scalebox{.4}{\includegraphics{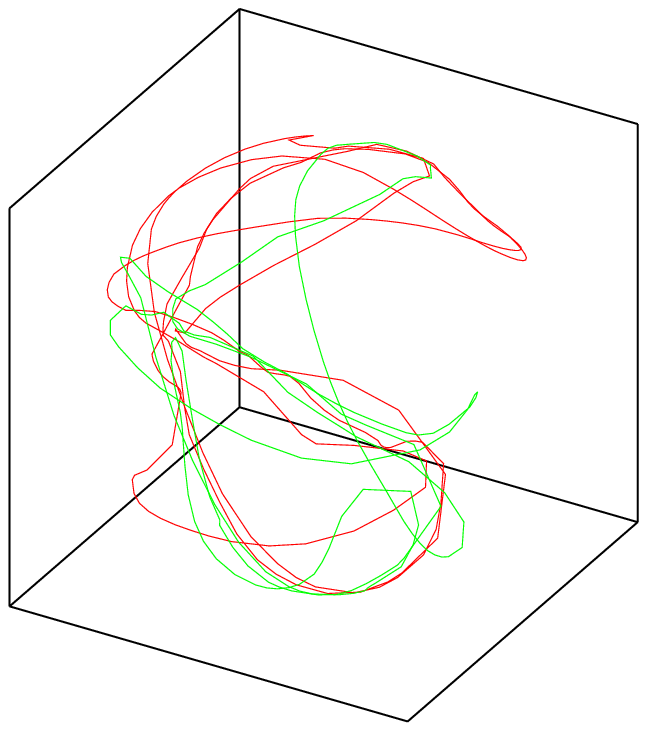}}
        & 
	\hspace*{-.3in}
        \scalebox{.4}{\includegraphics{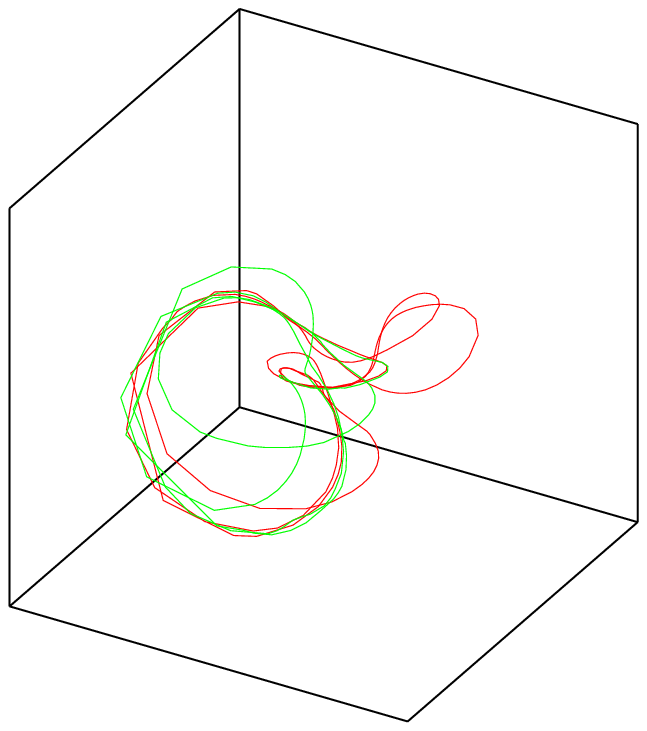}}\\
        (f) & (g) \vspace{-.2in}\\
	\hspace*{-.3in}
        \scalebox{.4}{\includegraphics{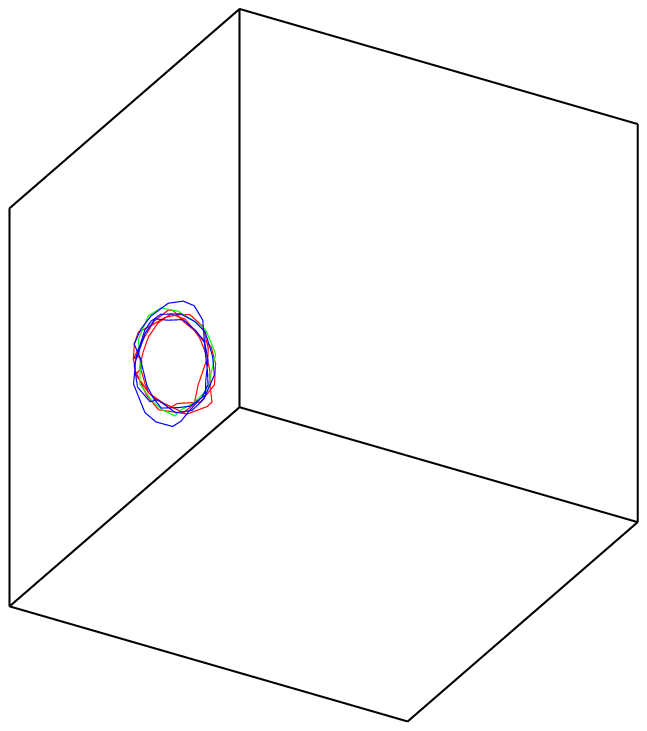}}
        & 
	\hspace*{-.3in}
        \scalebox{.4}{\includegraphics{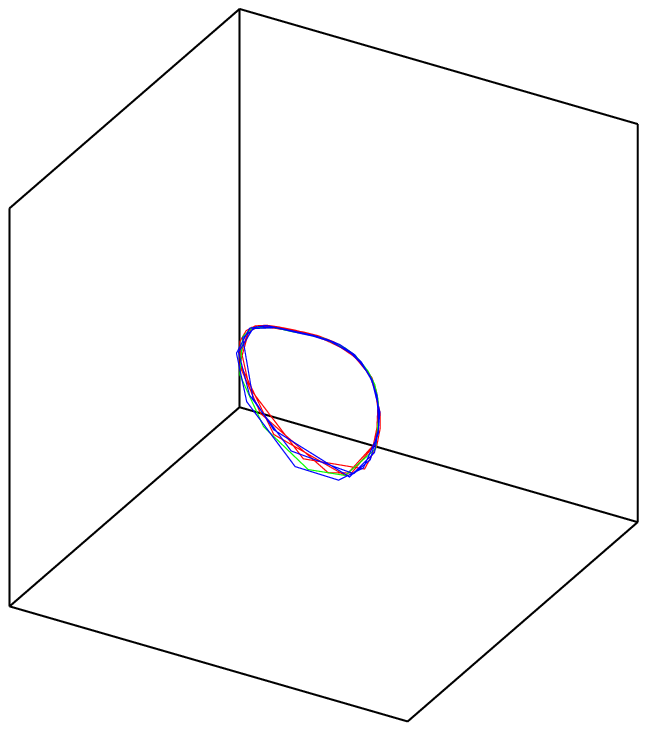}}	
        \end{tabular}
\caption{Development of vortices for $v_{ns}=1.4$ cm/s, with no
non-local contribution. Frame a shows the anisotropy density
as a function of time. Frames b, d, and f are Hopf projections
of the vortices at $t=0$, 0.5, and 1.6 s, all with the same scale.
Frames c, e, and g are stereographic projections at the same times,
again all with the same scale.}
\label{f:collapse}
\end{center}
\end{figure}

The flow lines of the Hopf field on the 3-sphere do not remain mutually
perpendicular throughout the entire space, so we observe nothing directly 
equivalent to the open-orbit state on the torus. However,
omitting the non-local interaction causes a different problem: a collapse of
vortex lines until they all lie nearly atop each other. In most cases the
vortices in the collapsed state follow roughly along a Hopf fiber, leading to an
extremely low anisotropy parameter. Figure \ref{f:collapse} illustrates the
collapse process at a driving velocity $v_{ns}=1.4$ cm/s. The first frame shows
the anisotropy parameter which initially varies between 0.1 and 0.2, then drops
abruptly to about 0.01, with little change thereafter. The remaining frames are
Hopf and stereographic projections at three times. The initial configuration is
a tangle equilibrated at the same driving velocity but with non-local terms
included. As seen from frames d) and e), upon omitting the non-local
contributions the vortex line density becomes less uniform, even before any
dramatic change in the anisotropy appears. The final frames f) and g) show the
complete collapse. Since the Hopf projection sends Hopf fibers to points, the
small size of the ring in f) indicates that the vortices stray little from a
single Hopf fiber. In frames f) and g) there are three distinct
vortices present, which make a combined total of seven circuits. Continuing the
calculation leads to trivial reconnections among these near-parallel vortices,
but no other interactions occur and the collapsed state persists indefinitely.

Low-velocity calculations without the non-local contribution routinely collapse to a set of nearly-overlapping loops, composing one or more vortex rings. While the loops are usually close to a Hopf fiber, with anisotropy parameter $I_{\parallel}<0.02$, that is not always
the case. One of our collapsed configurations had final anisotropy parameter an order of magnitude larger, $I_{\parallel}=0.27$. For our usual right-handed Hopf flow, the vortex loop is roughly aligned with the flow. Starting with what appears to be a stable configuration and inverting the direction of all vortices causes immediate vortex motion, as the loops rotate by $\pi$ relative to the Hopf flow. Similarly, for a left-handed Hopf flow stable vortices are anti-parallel to the field. This direction change is exactly as expected since vortices are axial vectors. Oddly, the persistence of such vortex arrangements depends on having several loops; upon removing all but a single loop, that loop moves and distorts until eventually additional loops are created. The requirement of multiple loops is particularly strange given that the LIA calculation does not include non-local interactions, so the loops interact solely through reconnections. The collapsed state appears to be a numerical artifact specific to using the local approximation on $S^3$, much as the open-orbit state is on $T^3$.

\begin{figure}[htb]
\begin{center}
        \begin{tabular}{l l}
         (a) & \hspace*{-.4in}(b) \vspace{-.2in}\\
	\hspace*{-.8in}
        \scalebox{.6}{\includegraphics{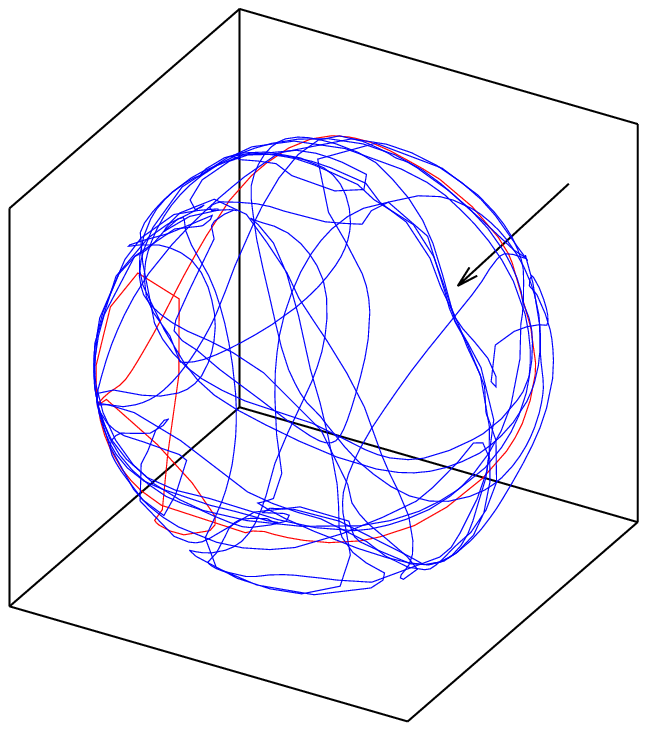}}
        & 
	\hspace*{-1.2in}
        \scalebox{.6}{\includegraphics{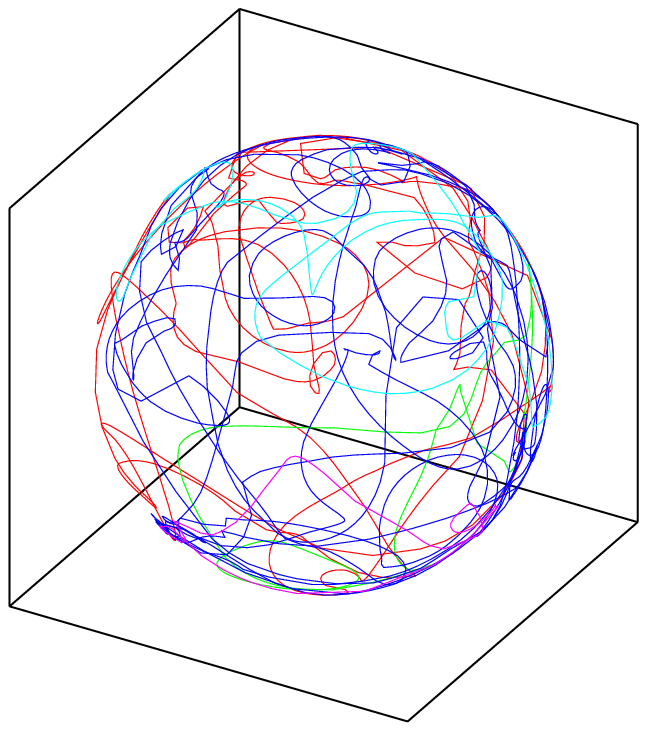}}	
        \end{tabular}
\caption{Hopf projections for tangles generated with no non-local contributions (a)
and with non-local contributions included out to a distance $d_{NL}=0.1\pi R$ (b).
For both tangles $v_{ns}=2$ cm/s and the sphere radius is $r=0.005$ cm. Vortex
clustering appears in (a), as indicated by the arrow, but is mostly eliminated in (b) by the short-distance non-local contributions.}
\label{f:smalldnl}
\end{center}
\end{figure}

For sufficiently large driving velocity, a non-trivial vortex tangle
persists even without non-local contributions, as illustrated in Figure
\ref{f:smalldnl}a. However, hints of the collapsed state remain, in the
tendency of vortices to clump together. Without the non-local term to
separate them, portions of vortices that become aligned will continue their
time development together until disrupted by reconnection with a non-aligned
vortex. As with the open-orbit state on $T^3$, adding non-local
contributions out to a very limited distance disrupts the collapsed state.
The tangle of Figure \ref{f:smalldnl}b, which results from including
contributions out to a distance $d_{NL}=0.1\pi R$, shows little sign of
the clustering that results from the purely local calculation. Our
calculations suggest that with non-local terms included even for such a
small fraction of the sphere, the tangles can persist indefinitely.

\begin{figure}[htb]
\begin{center}
\hspace*{-.15in}
\scalebox{.45}{\includegraphics{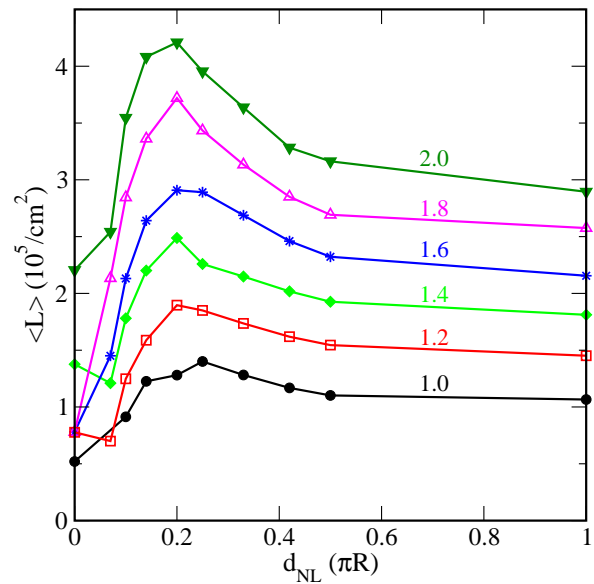}}
\caption{Equilibrated line length density as a function of the extent of the non-local calculation, for several applied velocity fields. The velocities, in cm/s, are indicated above each curve.}
\label{f:densitydnl}
\end{center}
\end{figure}

\section{Effect of non-local term}

\begin{figure}[bht]
\begin{center}
\hspace*{-.15in}
\scalebox{.45}{\includegraphics{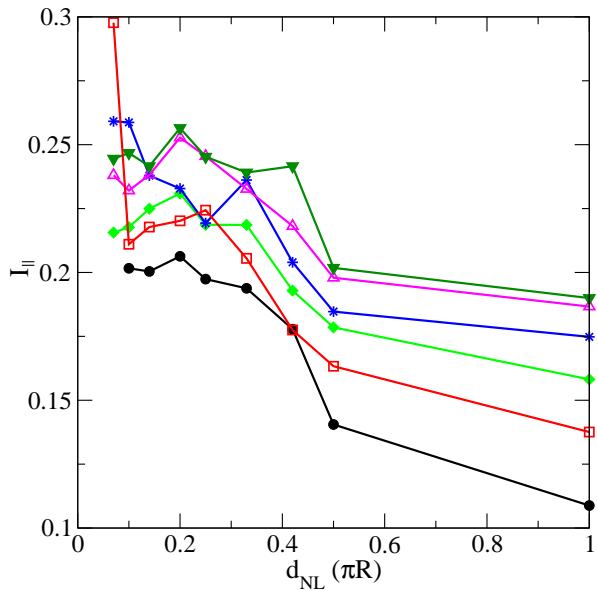}}
\caption{Anisotropy parameter as a function of the extent of the non-local calculation, for several applied velocities. The driving velocities, in cm/s, are 1.0, 1.2, 1.4, 1.6, 1.8, and 2.0, with $I_{\parallel}$ strictly increasing with drive at $d_{NL}=1$.}
\label{f:anisotdnl}
\end{center}
\end{figure}

\begin{figure*}[t]
\scalebox{.6}{\includegraphics{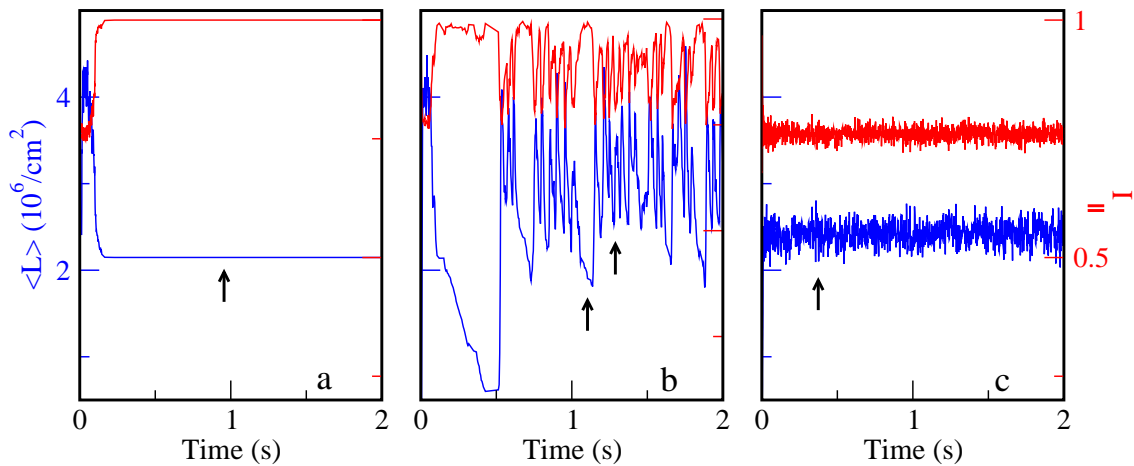}}
\caption{Line length density (lower curves, left axes) and anisotropy (upper curves, right axes) on a torus, with $v_{ns}=12$ cm/s.
a) LIA only, straight velocity field, b) LIA only, helical velocity field, c)
full Biot-Savart law, straight velocity field. Arrows indicate the configurations
shown in Figure \ref{f:confighelix}.}
\label{f:denshelix}
\end{figure*}

As with our previous simulations with the usual periodic boundary
conditions \cite{owentorus}, we examine the effects of including non-local
terms only for vortices within a distance $d_{NL}$. Figure
\ref{f:densitydnl} shows the results. At very small $d_{NL}$, the main
effect is to dissipate the clumping, since intervortex interactions
separate the vortex lines. Forcing the vortices apart also helps to
maintain a true tangle. Without the non-local interaction, the vortices can
collapse along a flow line, which prevents further energy gain or loss in
the tangle. The non-local term and resulting reconnections ensures that some
of the vortex loops have segments perpendicular to the velocity field. 
These perpendicular segments exchange energy with the applied field,
allowing the line density to grow. As $d_{NL}$ increases further, the line
length density of the tangle decreases gradually. This is analogous to our
previous results for periodic boundary conditions, where the non-local
interaction reduces the line length by favoring reconnections between
anti-parallel vortices over those between near-parallel vortices. However,
with periodic boundary conditions this decrease in $\langle L\rangle$ 
occurs mainly when $d_{NL}$ is smaller than the typical vortex separation.
At larger $d_{NL}$ the line length is nearly constant. Effectively, only
``nearest-neighbor" vortices contribute significantly to the velocity
field. The much more gradual effect of non-local terms in $S^3$ may arise
from the anisotropic nature of the tangles, which means that distant
non-local contributions are not directed randomly and do not cancel as
efficiently as they do on the torus. Interestingly, the crossover between
the low-$d_{NL}$ reduction of clumping and the high-$d_{NL}$ reduction of
line length occurs when the non-local term is applied over a distance
comparable to the vortex line spacing. In fact, the maximum line length
density seems to move towards smaller $d_{NL}$ as the applied velocity
field increases, consistent with the reduced line spacing at higher
velocities. 

As shown in Figure \ref{f:anisotdnl}, the tangles become slightly more
anisotropic as the driving velocity decreases, probably because at lower
line density there are fewer reconnections. The anisotropy is relatively
constant at low $d_{NL}$, but increases as the interactions extend beyond
one hemisphere. This increased anisotropy is entirely different from the
collapse observed as $d_{NL}\rightarrow 0$, and much more subtle. The
vortices remain intertwined, but with an incomplete tendency to align with
the velocity field. The increase in this alignment when the interactions
reach a full hemisphere suggests that the coupling of distant regions
through the full $S^3$ Biot-Savart law plays a significant role in the
anisotropy. In fact, extending the interactions from one hemisphere to the
entire sphere affects the anisotropy more than it does the line density. 
Figure \ref{f:anisotdnl} omits combinations of $d_{NL}$
and $v_{ns}$ for which the tangle collapses to a single trajectory of nearly
overlapping vortices. At the lowest $d_{NL}$ where the vortices do not
completely collapse, $I_\parallel$ relies heavily on a small number of
vortex loops and becomes erratic, for example at $v_{ns}=1.2$ cm/s and
$d_{NL} = 0.07\pi R$.   

\section{Polarization of tangles}

We now return to the significant difference in tangle anisotropy between
simulations on $T^3$ and $S^3$. In the torus calculations with
periodic boundary conditions we find a slight tendency of vortices to align
perpendicular to the driving velocity field. The Glaberson-Donnelly
instability \cite{Glaberson74}, which has been observed by various groups both experimentally \cite{Cheng73, Swanson83} and in
computational work \cite{Tsubota03, Mesgarnezhad18}, prevents parallel alignment.
Essentially, for a straight vortex parallel to an external velocity field,
certain distortions of the vortex away from straight tend to grow. With
multiple vortices, such growth ultimately leads to reconnections that alter
the vortex arrangement. The Glaberson-Donnelly instability also impacts
simulations using the local induction approximation, where an ``open-orbit"
state appears with non-interacting vortices aligned in a single direction.
The vortex direction is always perpendicular to the driving velocity because
of the instability.

By contrast, on $S^3$ the driving velocity field follows helical Hopf fibers.
This means that no set of parallel vortices, all perpendicular to the
velocity field throughout the 3-sphere, can exist. Another factor
weakening the Glaberson-Donnelly instability is that, unlike on the torus, the
longest-wavelength distortion of a flow line is itself another flow
line. This holds only for a distortion with handedness matching that of
the Hopf flow, which is consistent with our observation that vortex rings
can only settle stably in one direction relative to the Hopf flow. For
LIA calculations we find near-perfect alignment with the Hopf field. The
polarization drops once non-local interactions are included, although the
turbulence remains anisotropic with a tendency towards parallel alignment.

\begin{figure}[htb]
\begin{center}
        \begin{tabular}{l l}
        (a) & (b) \vspace{-.2in}\\
	\hspace*{-.3in}
        \scalebox{.4}{\includegraphics{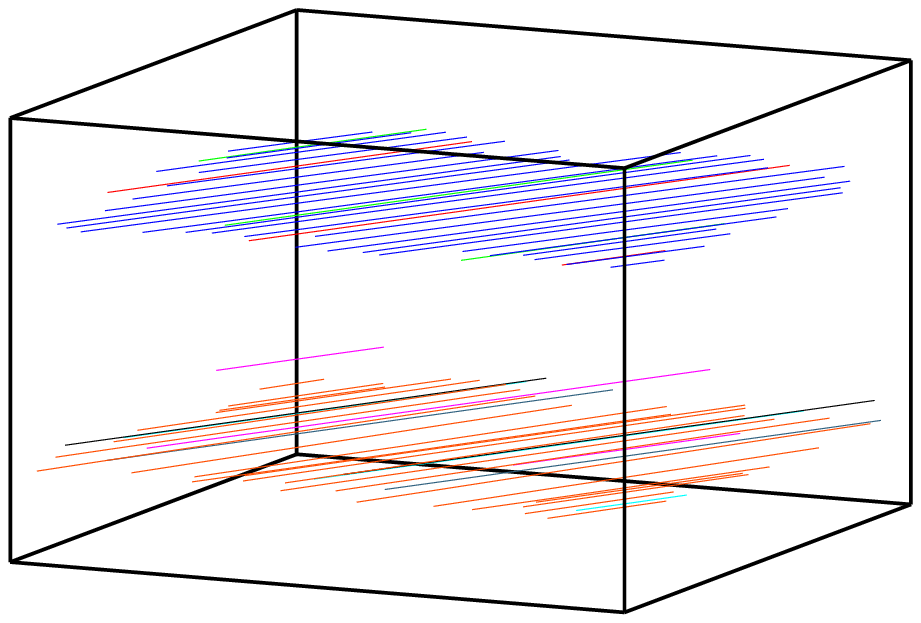}}
        & 
	\hspace*{-.3in}
        \scalebox{.4}{\includegraphics{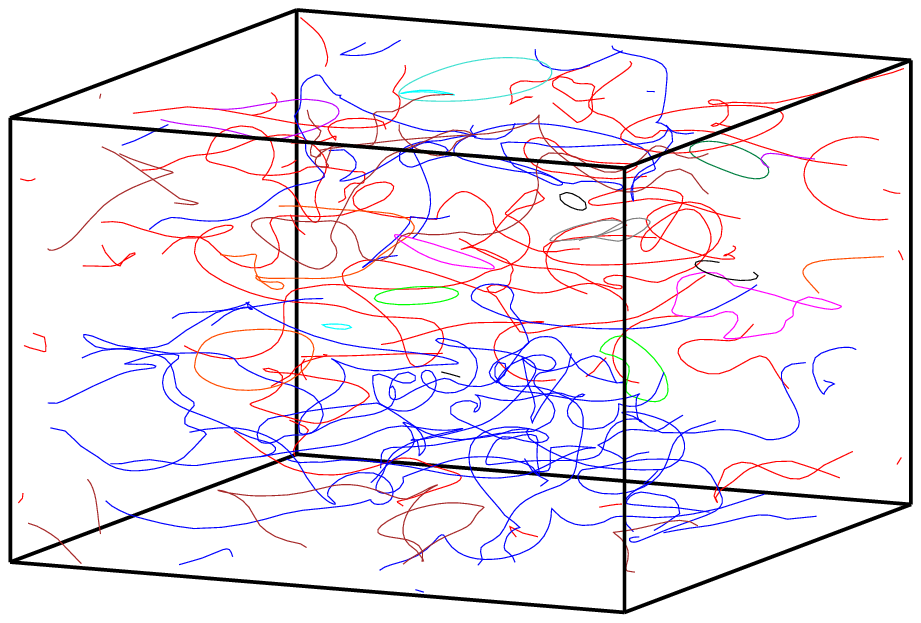}}\\
        (c) &(d) \vspace{-.2in}\\
	\hspace*{-.3in}
        \scalebox{.4}{\includegraphics{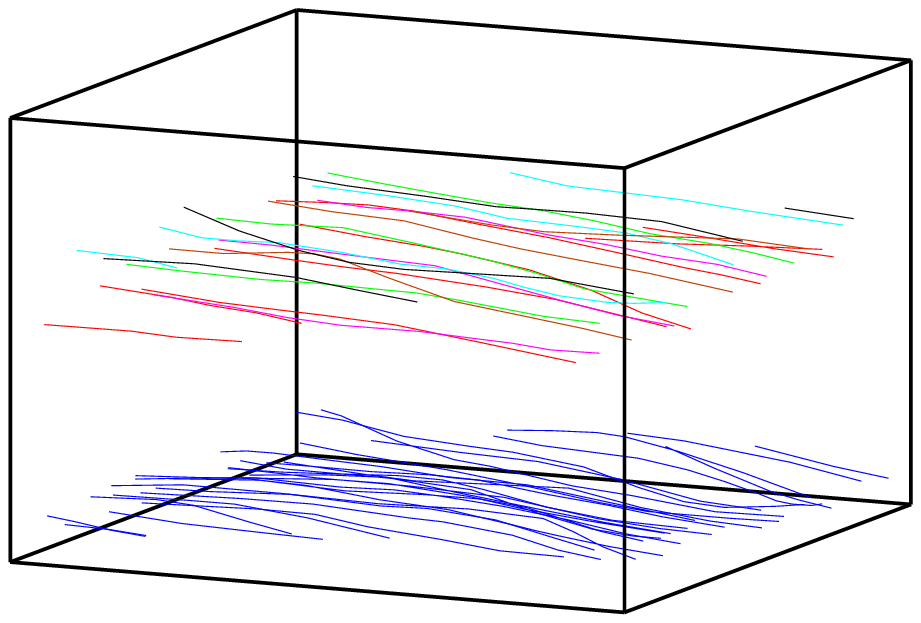}}
        & 
	\hspace*{-.3in}
        \scalebox{.4}{\includegraphics{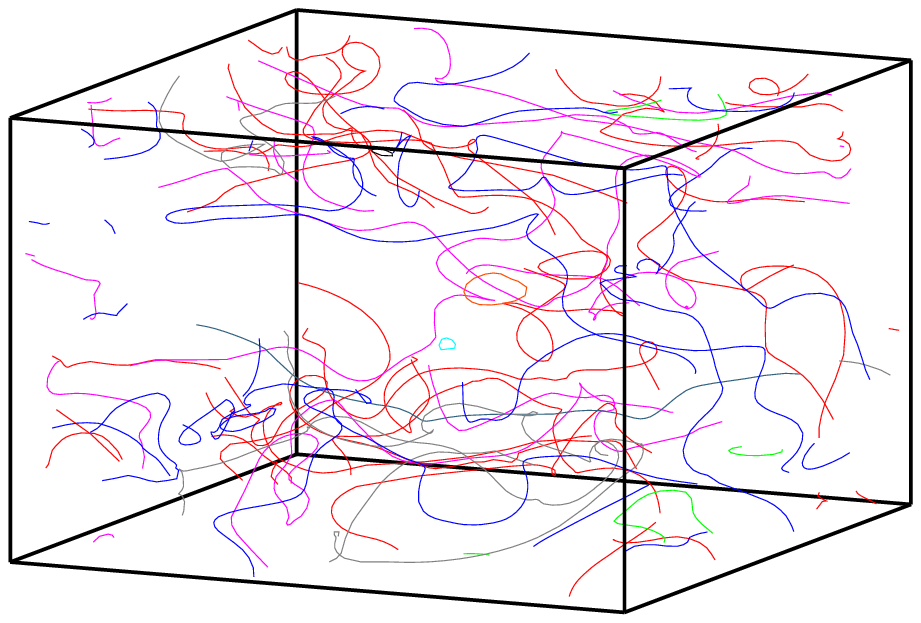}}
        \end{tabular}
\end{center}
\caption{Vortex configurations for torus simulations, corresponding to
the points noted in Figure \ref{f:denshelix}.
a) LIA only, straight velocity field, b) full Biot-Savart law,
c) LIA only, helical velocity field, high anisotropy, d) LIA only,
helical velocity field, low anisotropy.}
\label{f:confighelix}
\end{figure}

This raises the question of whether a helical velocity field on a
torus would lead to some of the same behavior as the Hopf flow on the
3-sphere. We carried out a few such simulations, using a helical
field around the $z$-axis. The field makes three rotations and has horizontal magnitude 10\% of the total. The helical
field indeed prevents the full open-orbit state from developing even when using
the local induction approximation. However, the vortex line density still
differs from that of the full non-local calculation. Figure \ref{f:denshelix}
shows the relatively stable line density from a full non-local calculation,
along with the perfectly stable line density from LIA once the open orbit
state has formed. The exact open-orbit line density varies depending on the initial conditions; in this case it is slightly lower than that of the non-local
calculation. The helical velocity field leads to entirely different behavior. The tangle undergoes frequent excursions towards the open-orbit state, with the anisotropy coefficient becoming very large; Figure \ref{f:confighelix}c illustrates a configuration with $I_\parallel=0.992$ and $\langle L\rangle=1.9\times 10^6$/cm$^2$. Unlike in the complete open-orbit state of Figure \ref{f:confighelix}a, slight curvature and misalignment remain. Remarkably, the helical velocity field enables recovery even from such a well-aligned configuration. A short time later, the vortices reach the configuration of Figure \ref{f:confighelix}d, with $I_\parallel=0.787$ and $\langle L\rangle=3.4\times 10^6$/cm$^2$. The tangle remains noticeably less homogeneous than that of Figure \ref{f:confighelix}b, which comes from a fully
non-local calculation and has $I_\parallel=0.787$ and $\langle L\rangle=2.6\times 10^6$/cm$^2$. Figure \ref{f:denshelix}b shows repeated fluctuations between these limits. The dependence on the helicity of the applied velocity field is entirely an effect of LIA; a full non-local calculation with helical applied velocity has density and anisotropy unchanged form Figure \ref{f:denshelix}c.

\section{Conclusions}

We have carried out vortex simulations on a 3-sphere. Our results show the potential importance of the global geometry and topology used. The most basic features remain the same as in standard torus-based calculations. Vortex tangles appear, with the vortex line length stabilizing at a value that increases with the driving velocity. However, the tangles are much less isotropic than those generated on a torus. Exponents relating the line length to the driving velocity
and the reconnection rate reflect this increased anisotropy. This suggests the intriguing possibility of generating steady-state tangles with markedly different properties by changing the underlying global structure. More broadly, computations throughout condensed matter are done on a torus, often without investigating what features may stem from the topology itself.

In calculations using the local induction approximation, the influence of the manifold was even more striking. The particular failure mode on $T^3$, the open orbit state, did not appear, but vortices instead collapsed to occupy a thin tube, all with nearly parallel alignment. A helical driving velocity on $T^3$, which shares some properties with the flow on $S^3$, avoids the full open orbit state under LIA but retains much of the same character. This is further evidence of the role of the manifold rather than the specific velocity field.

\begin{center}
\textbf{ACKNOWLEDGMENTS}
\end{center}

We thank G. Kuperberg for many helpful conversations, including bring to our attention the exact Biot-Savart formula on $S^3$. One of us (Dix) acknowledges funding from a Department of Education GAANN fellowship.\\

\begin{center}
\textbf{APPENDIX}
\end{center}

\noindent
{\em Stereographic Projection}\\

\begin{figure}[htb]
\begin{center}
\scalebox{.5}{\includegraphics{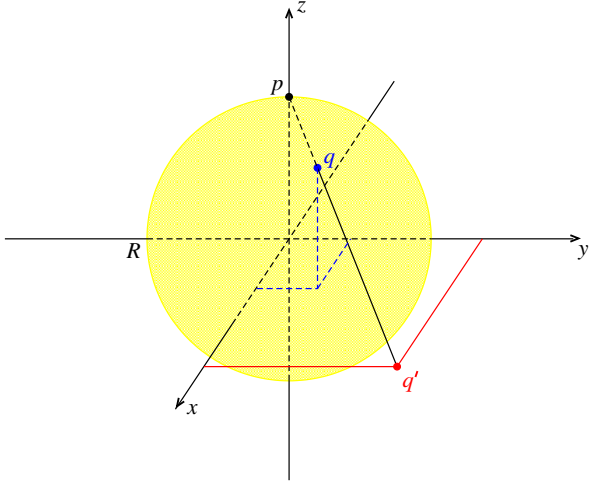}}
\caption{Illustration of stereographic projection from the surface of
a three-dimensional ball to the $xy$ plane. A point $q$ maps to $q^\prime$
in the plane such that $q$, $q^\prime$, and $p$ are collinear.}
\label{f:stereo3d}
\end{center}
\end{figure}

In this case $S^3$ is mapped onto $\mathbb{R}^3$. Figure \ref{f:stereo3d}
shows the analogous operation mapping $S^2$ to the $z=0$ plane. The
projection point $p$ is (0,0,$R$). A line is extended from $p$ through
a point on the circle, and the intersection between that line and the
$xy$-plane gives the image point. Under this projection, the equator maps
to a circle of radius $R$. The lower hemisphere goes to the inside of
this circle, and the outer hemisphere to the outside. The projection point
itself does not have an image within the plane, but maps to a point
at infinity. In general, the image of a point $(x,y,z)$ in $S^2$ has
coordinates 
$$(\frac{x}{1-z/R},\frac{y}{1-z/R})$$
in the plane. The generalization to a projection from $S^3$ to, say,
the $w=0$ space is straightforward.

The main disadvantage of stereographic projection is that distances are
not preserved. One half of $S^3$ maps to the inside of a finite sphere,
while the other half maps to the infinite outside. This can give a
misleading impression of the structure of a vortex tangle. \\

\noindent
{\em Hopf Fibration}\\

A Hopf fibration maps $S^3$ onto $S^2$ in a way that preserves
distance, up to an overall scale factor. It is selected to match
our driving velocity field, $\mathbf{v}=\frac vR(-y,x,-w,z)$; as we show
below, the corresponding Hopf fibration maps each flow line to a single
point. We take the Hopf fibration as the following map to $\mathbb{R}^3$:
\begin{center}
$h_1=2(xz+yw)$\\
$h_2=2(yz-xw)$\\
$h_3=(x^2+y^2)-(z^2+w^2)$
\end{center}
Explicit multiplication shows that for any point on a 3-sphere of radius
$R$, its image is at radius $R^2$. Thus $S^3$ in fact maps to $S^2$.

The flow lines of the velocity field are particular great circles
on $S^3$.  To track them under the Hopf fibration, consider the
four-dimensional real space which contains $S^3$ as a two-dimensional
complex space with coordinates $q_1=x+iy$ and $q_2=z+iw$. A complex line,
$q_2=kq_1$, intersects $S^3$ in a great circle of constant $|q_1|$ and
$|q_2|$, which is exactly a flow line of this velocity field. In addition,
writing the Hopf fibration in terms of the complex coordinates and using
$q_2=kq_1$ gives
\begin{center}
$h_1=Re(2q_1q_2^*)=2|q_1|^2Re(k)$\\
$h_2=Im(2q_1q_2^*)=2|q_1|^2Im(k)$\\
$h_3=|q_1|^2-|q_2|^2=|q_1|^2(1-|k|^2)$
\end{center}
The image depends only on $k$ and $|q_1|$, both of which are constant
on any great circle corresponding to a flow line of the velocity field.
Hence this Hopf fibration maps a flow line on $S^3$ to a point on $S^2$.

Any two Hopf fibers of the flow are great circles that twist around
each other. The distance between any point on one fiber and the closest
point on the other is constant, which gives a sensible definition for
the distance between the two fibers. The Hopf fibration preserves this
distance, up to an overall scale factor. For example all fibers a fixed
distance from a given fiber $A$ are mapped to a circle on $S^2$ centered
at the image of $A$.

\end{document}